\documentclass[aps,prl,twocolumn,showpacs,superscriptaddress]{revtex4}
\usepackage{bm}
\usepackage{color,soul}
\usepackage{epsfig}
\def\Kh{h_0}
\def\lm{\lambda}
\begin{document}

\title{Stripe formation in bacterial systems with density-suppressed
  motility}
\author{Xiongfei Fu}
\affiliation{Department of Physics, The University of Hong Kong,
  Pokfulam, Hong Kong, China}
\author{Lei-Han Tang}
\affiliation{Department of Physics, Hong Kong Baptist University, Kowloon Tong, Kowloon, Hong Kong, China}
\affiliation{Beijing Computational Science Research Center, 3 Heqing Road, Haidian,  Beijing 100084, China}
\author{Chenli Liu}
\affiliation{Department of Biochemistry, The University of Hong Kong,
  Pokfulam, Hong Kong, China}
\author{Jian-Dong Huang}
\affiliation{Department of Biochemistry, The University of Hong Kong,
  Pokfulam, Hong Kong, China}
\author{Terence Hwa}
\affiliation{Center for Theoretical Biological Physics, University of California at San Diego, La Jolla, CA, USA}
\author{Peter Lenz}
\affiliation{Department of Physics and Center for Synthetic Microbiology, University of Marburg, Marburg, Germany}
\date{\today }

\begin{abstract}
  Engineered bacteria in which motility is reduced by local cell
  density generate periodic stripes of high and low density when
  spotted on agar plates. We study theoretically the origin and
  mechanism of this process in a kinetic model that
  includes growth and density-suppressed motility of the cells.  The
  spreading of a region of immotile cells into an initially
  cell-free region is analyzed. From the calculated front profile we
  provide an analytic ansatz to determine the phase boundary between
  the stripe and the no-stripe phases. The influence of various
  parameters on the phase boundary is discussed.
%{\bf (590 characters (including spaces) with 600 being allowed.)}
\end{abstract}

\maketitle

Biological systems exhibit a wide variety of exquisite spatial and
temporal patterns. These patterns often play vital roles in
embryogenesis and development \cite{chuo09,kond10}. In addition,
colonies of bacteria and simple eukaryotes also generate complex
shapes and patterns
\cite{ben00,budr91,fuji92,mats00,pedl92,welc01}. Typically, these
patterns are the outcome of coordinated cell growth, movement, and
differentiation that involve the detection and processing of
extracellular cues \cite{ben00}.

These experimental observations have triggered extensive mathematical
modeling. A large body of theoretical work is devoted to pattern
formation by chemotactic bacteria.  On the mean-field level, these
phenomena can be described by Keller-Segel type reaction-diffusion
models \cite{kell70,murray,schn93}. In many instances, the models
invoke non-linear diffusion of the cells where the diffusion
coefficient increases with the local cell density \cite{ben00,kawa97}.

\begin{figure}[b]
  \centering
   \includegraphics[width=.45\textwidth,clip]{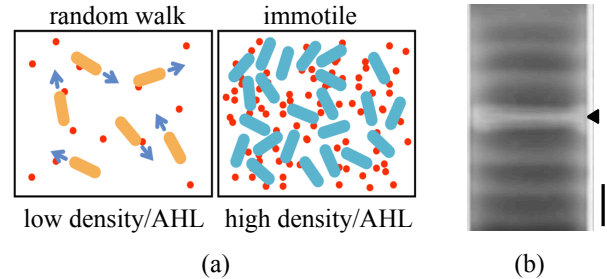}
   \caption{ (a) The engineered bacterium cells execute ``random
     walks'' at low densities but become immotile at high
     densities. (b) This coupling between density and motility leads
     to the formation of stripes with periodic density variations on
     agar plates \cite{exp_paper}. Initial cell seeding was done (at
     the position indicated by the arrow) 30hr before the picture was
     taken. Bar corresponds to a length of 5mm. \label{fig:c1}}
\end{figure}

Recently, it was theoretically proposed that the opposite case of
density suppressing motility could also lead to patterns via a
``self-trapping'' mechanism \cite{tail08,cate10}. In parallel, we have
explored such a system experimentally, using a synthetic biology
approach \cite{exp_paper}. The density-suppressed motility was
introduced into the bacterium {\em E.~coli} by having it excrete a
small (and rapidly degraded) signaling molecule AHL, such that at low
AHL levels, these cells perform random walks via their swim-and-tumble
motion \cite{berg04} and are ``motile", while at high AHL levels,
these cells tumble incessantly, resulting in a vanishing macroscopic
motility and becoming ``immotile" [Fig.~\ref{fig:c1}(a)].

On agar plates, these engineered bacteria form highly regular and
stable stripe patterns consisting of periodically alternating regions
of high and low cellular densities [Fig.~\ref{fig:c1}(b)]. A thorough
characterization of these spatial patterns gave rise to the following
key experimental observations \cite{exp_paper}: (i) Regulation of cell
motility by AHL is essential for pattern formation; (ii) Cells are
motile at low densities and immotile at high densities; (iii)
Bacteria form stripes sequentially in one- and two-dimensional geometries
when expanding into an initially cell-free region; 
(iv) Random initial conditions do not give rise to stripes; (v) Chemotaxis
is not required for pattern formation; (vi) The stripe patterns depend
on the magnitude of the unrepressed cellular motility in the low
density limit: Upon decreasing this magnitude the system makes a
transition from a phase with spatially periodic stripes (the stripe
phase) to the no-stripe phase, through a region with a finite number
of stripes.

As demonstrated in~\cite{exp_paper}, all the experimental observations
can be reproduced by a three-component model that (i) describes the
cellular motion as random walk with an abrupt AHL-dependent motility
coefficient, (ii) takes into account the synthesis, diffusion, and
turnover of AHL, and (iii) implements the consumption and diffusion of
the nutrient due to cell growth and the limitation of growth in the
absence of nutrient.

Despite the success of this model, the origin and mechanism of the
pattern formation process remain unclear. In this paper, we 
describe a simplified two-component model to study the essential features
of stripe formation analytically. In terms of the concentration $h(x,t)$ of AHL
and the cell density $\rho(x,t)$ at position $x$ and time $t$,
the dynamical equations are given by,
\begin{eqnarray}
\frac{\partial h}{\partial t} & = & D_h\frac{\partial^2h}{\partial
x^2}+\alpha\rho-\beta h ,\label{eq:1}\\
\frac{\partial \rho}{\partial t} & = & \frac{\partial^2}{\partial
x^2}[\mu(h)\rho]+\gamma \rho \left (1-\frac{\rho}{\rho_s} \right).\label{eq:2}
\end{eqnarray}
The first equation describes production (with rate $\alpha$),
diffusion (with diffusion coefficient $D_h$) and turnover (with rate
$\beta$) of AHL.  It is clear that in spatially homogeneous
situations, $h \propto \rho$ in the steady state, hence the name
``quorum sensor'' for AHL. The second term on the right hand side of
Eq.~(\ref{eq:2}) describes logistic bacterial growth at rate $\gamma$
and with a saturation density $\rho_s$. The reduced growth rate at
high densities approximates the nutrient depletion effect in the
experiments.  The stochastic swim-and-tumble motion of the bacteria is
modeled as a diffusion-like term on the right hand side of
Eq.~(\ref{eq:2}). The experimentally measured values of all parameters
can be found in Ref. \cite{exp_paper}.

The motility function $\mu(h)$ explicitly depends on $h$. It takes
into account the repressive effect of AHL concentration (and hence
cell density) on cell motility.  The interaction term in
Eq.~(\ref{eq:2}) can be obtained by either generalizing the
coarse-graining procedure of Ref. \cite{tail08} or adopting the master
equation approach of Ref. \cite{murt78} to an $h$-dependent motility.
In fact, such an analysis yields a mixture of two terms $\partial _x
(\mu(h)\partial _x \rho)$ and $\partial_x (\rho \partial _x\mu(h))$
(for details see SM). For simplicity we focus on the above coupling, 
%\footnote{
%In a well-developed colony of the engineered bacteria where
%the AHL concentration $h(x,t)$ varies smoothly in space, regions of low and high
%motility can be identified. The front at $h\simeq\Kh$ is sharp
%due to the observed step-like diffusion constant reported in Ref.~\cite{exp_paper}.
%Bacteria within the interface region exhibit reduced diffusivity of varying degree,
%which is difficult to describe precisely. Nevertheless, it is clear that the interface 
%acts as an absorbing boundary for the diffusing cells on the motile side with 
%a very low escaping probability. It is in this sense that the diffusion term in Eq.~(\ref{eq:2})
%is to be understood.}
but our main conclusions are not affected by this (for details see SM).

Measurements of bacterial diffusion at the population level show that
$\mu$ drops abruptly from a value $D_{\rho}$ to $D_{\rho,0}\ll D_\rho$ 
as $h$ increases beyond a threshold $\Kh$. 
As simulation results of Ref. {\cite{exp_paper}} did not
depend sensitively on the value $D_{\rho,0}$, we shall set $D_{\rho,0}=0$.  
Thus, we consider the form $\mu(h)=D_{\rho}$ for $h \leq \Kh-w$ and $\mu(h)
=0$ for $h> \Kh$ with a linear decrease of $\mu$ for the transition
region $\Kh-w <h<\Kh$ with $\Kh \gg w \rightarrow 0$.

As demonstrated in Ref.~\cite{exp_paper}, this two-component model is
able to initiate stripe patterns in a growing bacteria colony and maintain them for a while; 
but the stripes are eventually lost after long times when cell densities
reach $\rho_s$ throughout the system. The latter behavior deviates from the 
experimental system where stripes are frozen in upon nutrient exhaustion. 
Nevertheless, the model correctly captures the dynamics at the
propagating front where new stripes are formed.
The simplicity gained enables analytic treatment that clarifies conditions 
for spontaneous stripe formation in the system. 

Consider a one-dimensional bacterial colony development as depicted 
in Fig.~\ref{fig:c1}(b). 
%From Eqs.~(\ref{eq:1})-(\ref{eq:2}), it is already apparent
%how regions of high cell density can form: 
Initially, cell density is low on
the plate and all cells grow and freely diffuse. As growth proceeds cells at the center
aggregate. The increased cell population boosts the local AHL concentration,
driving it eventually above $\Kh$ so that cells inside the aggregate become
immotile. At the same time, this high density
region expands outward by absorbing cells moving from surrounding
low-density regions into the aggregate.
Depending on the parameter values of the system, the high density
region expands either stably as a front or exhibits instability that
results in stripes \cite{exp_paper}. 

We now take a closer look at the low-density region that precedes the
advancing aggregate, whose cell density profile is calculated later
(see Fig.~\ref{fig:1}). The size of this motile cell population is
maintained by a dynamic balance between cell growth within and loss to
the aggregate in the contact zone. Due to absorption by the aggregate,
cell number is low in the contact region.  By virtue of
Eq.~({\ref{eq:2}), the maximum density $\rho_m$ of motile cells is
  found at a distance $L_\rho=\sqrt{D_\rho/\gamma}$ from the
  aggregate, while the cell diffusion flux into the aggregate is given
  by $J\simeq D_\rho\rho_m/L_\rho$.  Meanwhile, the expansion speed
  $c$ of the aggregate satisfies $J=c\rho_c$ where $\rho_c$ is the
  density drop across the aggregate boundary.  Hence quite generally
  $\rho_m\simeq\rho_c$, i.e., the cell density profile in the motile
  region scales with the density at the edge of the aggregate where
  the AHL concentration is at the threshold value $\Kh$.  A
  quantitative calculation is then required to determine whether the
  AHL concentration rises to the threshold again at $\rho_m$.  As we
  shall see below, the answer depends on how the diffusion length
  $L_h=\sqrt{D_h/\beta}$ of AHL molecules (i.e., the typical distance
  travelled by an AHL molecule before degradation) compares to
  $L_\rho$.

%When the colony has grown to sufficient size, we may just focus on one side
%of the system. Let $x$ be the distance from the center line where growth
%was initiated. In stable growth, Eqs.~(\ref{eq:1})-(\ref{eq:2}) admit traveling
%wave solutions $h(x,t)=h(x-ct)$ and $\rho(x,t)=\rho(x-ct)$. 
%Equations~(\ref{eq:1}) and (\ref{eq:2}) are then reduced to two simple ordinary 
%differential equations taking on two different forms in respective regions. 
%From the solution 
%(given by the solutions of the equations in the
%two regions with matching boundary conditions at $z=0$), 
%we can check the validity of our assumption that $h$ crosses $\Kh$ only once. 
%For parameters where the solution is not self-consistent, the simple
%moving front is unstable, yielding an estimate of the phase boundary.

We will analyze a rescaled version of the
model~(\ref{eq:1})-(\ref{eq:2}) that only depends on dimensionless
quantities. We measure length in units of $L_\rho$, time in units of
$1/\gamma$, $\rho$ in units of $\beta \Kh/\alpha$ and $h$ in units of
$\Kh$. All dimensionless quantities are denoted by a hat (e.g. $\hat t
\equiv t\gamma$ etc.).  For a steadily propagating front at speed
$\hat{c}$, the density profiles $\hat{\rho}$ and $\hat{h}$ become
functions of $\hat{z}=\hat{x}-\hat{c}\hat{t}$. We set the front
position at $\hat{z}=0$ such that cells are immotile for $\hat{z}<0$
(region I) and motile for $\hat{z}>0$ (region II).

The cell density profile in region I is easily obtained by
integrating Eq.~(\ref{eq:2}) with the boundary condition $\hat
\rho_{\rm I}(-\infty)=\hat \rho_s\equiv \alpha\rho_s/(\beta \Kh)$,
\begin{equation}
\hat \rho_{\rm I}(\hat{z})=\frac{\hat{\rho_s}
\hat{\rho_c}}{\hat{\rho_c}+(\hat{\rho_s}-\hat{\rho_c})e^{\hat{z}/\hat c}}, 
\label{eq:6}
\end{equation}
where $\hat \rho_c\equiv \hat \rho_{\rm I}(0^-)$ is the scaled cell density at 
the edge of the aggregate. The experimental system of Ref.~\cite{exp_paper}
has a $\hat \rho_s\simeq 4$.

In region II, the marginal stability criterion \cite{saar88} yields
$\hat \rho_{\rm II}(\hat z)\sim e^{-\hat z}$ for $\hat z \gg 1$ with
the selected wave speed $\hat c=2$. With this choice, Eq.~(\ref{eq:2})
takes on the following form (except within a distance $w$ from the
interface),
\begin{equation}
\label{eq:4_new}
\hat\rho_{\rm II}''+2 \hat\rho_{\rm II}'+\hat\rho_{\rm II} \left (1-\hat\rho_{\rm II}/\hat\rho_s \right )=0,
\end{equation}
where the prime denotes $d/d\hat z$.  For the form of $\mu(h)$
described, one finds $\hat \rho_{\rm II}(0^+)=0$ as $w \rightarrow 0$
(see SM). Matching the diffusional flux from the motile side with the
speed of the immotile front yields the second condition at the
interface: $\hat \rho_{\rm II}'(\hat z=0^+)=2 \hat \rho_c$. Thus,
$\hat \rho_{\rm II}(\hat z)$ is a non-monotonic function, rising for
small $\hat z$ before decaying exponentially for large $\hat z$.

The AHL profile is determined from the cell density profile as
\begin{equation}
\label{eq:6_new}
\hat h(\hat z)=\hat \beta\int_{-\infty}^\infty d\hat z_1\hat
\rho(\hat z_1)G_h(\hat z-\hat z_1),
\label{h_0}
\end{equation}
with the Green's function,
\begin{equation}
G_h(\hat z)=-(1+\hat D_h\lm)e^{-\hat z/\hat D_h}e^{(1+\hat D_h\lm)|\hat z|
/\hat D_h}/2,
\label{G_h}
\end{equation}
and $\lm \equiv - [1 + (1+\hat D_h \hat \beta)^{1/2}]/\hat D_h$.
%and $\lm \equiv - (1 + \sqrt{1+\hat D_h \hat \beta})/\hat D_h$.

%We now describe the qualitative features of the AHL profile that follow
%directly from Eqs.~(\ref{h_0}) and (\ref{G_h}) without full solution of the problem.
%Deep inside the aggregate, both $\hat \rho(\hat z)=\hat\rho_s$ and 
%$\hat h(\hat z)=\hat \rho_s$ assume constant values.
%Obviously we need $\hat\rho_s>1$ for consistency.  
%Far in front of the front, the cell density and hence
%$\hat h(\hat z)$ decays to zero exponentially provided
%$|\lm|>1$. (Otherwise, the AHL wave will travel at faster speed
%than $\hat{c}$, invalidating the one-wave ansatz introduced above.)  In
%between the two limits, how closely $\hat h(\hat z)$ follows the
%non-monotonic cell density profile $\hat\rho(\hat z)$ depends on the
%two parameters $\hat D_h$ and $\hat \beta$ that specify the range of
%averaging in Eq.~(\ref{h_0}).  

Due to the nonlinearity in Eq.~(\ref{eq:4_new}), an exact solution for
$\hat \rho_{\rm II}$ is not possible and we shall analyze the problem
in an expansion in $\varepsilon\equiv \hat \rho_c/\hat \rho_s=\rho_c/\rho_s$. We
shall first consider the limit $\varepsilon\rightarrow 0$.  The
solution~(\ref{eq:6}) is then approximated by $\hat\rho_{\rm I}(\hat
z)\simeq \hat\rho_{\rm I}^{\rm lin}(\hat z)=\hat \rho_c e^{-\hat
  z/2}$.  The linear form of Eq.~(\ref{eq:4_new}) together with the
matching conditions at $\hat z=0$ yields $\hat \rho_{\rm II}^{\rm
  lin}(z)=2 \hat \rho_c \hat z e^{-\hat z}$.  Inserting $\hat\rho^{\rm
  lin}(\hat z)$ into (\ref{eq:6_new}), we obtain the AHL profile to
the zeroth order in $\varepsilon$,
\begin{equation}
\label{h-linear}
\hat  h ^{\rm lin}_{\rm II}(\hat z)=\hat{\beta} \hat{\rho_c}\Bigl(\frac{
4-4\hat D_h}{v^2}  e^{-\hat z}
+\frac{2}{v} \hat ze^{-\hat z}
+\frac{\lm^2}{w}e^{\lm\hat z}\Bigr),
\end{equation}
with $v \equiv 2-\hat D_h+\hat \beta$ and $w \equiv
(1+\lm)^2(1+2 \lm) (1+\hat D_h \lm)$. The value of
$\hat \rho_c$ is determined by the definition of the front at
$\hat z=0$, i.e. $\hat h ^{\rm lin}_{\rm II}(0)=1$.

Higher order corrections to the analytical profiles given above can
be computed systematically by rewriting Eq.~(\ref{eq:4_new}) in the
form,
\begin{equation}
\hat\rho_{\rm II}(\hat z)=\hat \rho_{\rm II}^{\rm lin}(\hat
z)+\frac{1}{\hat\rho_s}\int_0^\infty d\hat z_1 \hat \rho_{\rm II}^2(\hat
z_1)G_\rho^{\rm lin}(\hat z-\hat z_1),
\label{rho_II_green_sol}
\end{equation}
where $G_\rho^{\rm lin}(\hat z)= \hat z e^{-\hat z} \theta(\hat z)$
(with $\theta(x)$ denoting the Heaviside function) is the Green's
function for the linear part of Eq.~(\ref{eq:4_new}).
Iteration of Eq.~(\ref{rho_II_green_sol}) yields $\hat \rho_{\rm
  II}(\hat z)$ as a power series in $\varepsilon$. 
% Eq.~(\ref{rho_II_green_sol}) can be used iteratively to obtain
% $\hat \rho_{\rm II}(\hat z)$ as a power series in $\varepsilon$. 
The result, together with $\hat \rho_{\rm I}(\hat z)$
given by Eq.~(\ref{eq:6}), can then be fed into Eq.~(\ref{h_0}) to give
$\hat h(\hat z)$.

\begin{figure}[t]
  \centering
  \includegraphics[width=.43\textwidth,clip]{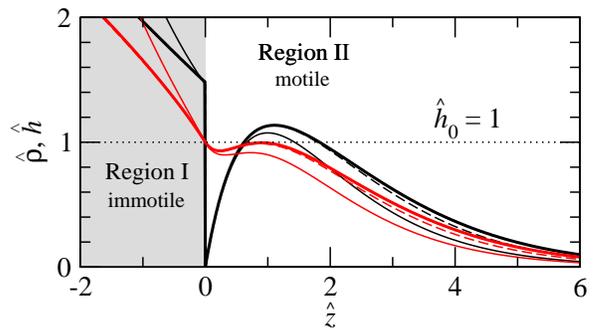}
  \caption{ (Color online.) Profiles of scaled cell density
    $\hat\rho(\hat z)$ (black) and AHL concentration $\hat h(\hat z)$
    (red) around the edge of the advancing aggregate at $\hat z=0$. Shown here are the
    analytical solution to the zeroth order (thin solid lines) and first order (dashed lines)
    in $\varepsilon=\hat\rho_c/\hat\rho_s$, and the numerically exact solution to the steady
    traveling-wave equations (thick solid lines). Here  $\hat D_h =
    D_h/D_\rho=1$,  
    $\hat \beta=\beta/\gamma=4$, and $\hat \rho_s=\alpha\rho_s/\beta \Kh=4$. 
    \label{fig:1}}
\end{figure}

We have carried out the above procedure to the first order in
$\varepsilon$.  Figure 2 shows typical $\hat h$- and $\hat
\rho$-profiles as obtained from our zeroth order (thin solid lines)
and first order (dashed lines) analytical solution for $\hat \rho_s=4$.
%(the parameter value for the experimental system of Ref. \cite{exp_paper}).  
As anticipated earlier, the $\hat\rho$-profiles
(black) for the motile population have the shape of a bulge with a depletion 
zone right ahead of the front at $\hat z=0$. 
%This is a consequence of the
%absorbing boundary condition at $\hat z=0$ (i.e. $\hat \rho_{\rm II}(0^+)=0$). 
In the zeroth order approximation, the bulge is
located at $\hat z=1$ with a peak value $\hat \rho_m^{\rm lin}=2\hat
\rho_c/e\simeq 0.736\hat\rho_c$.  For the values of $\hat D_h$ and
$\hat \beta$ shown, the AHL profiles (red) also develop a dip in
the contact zone. Nonetheless, the traveling wave solutions are self-consistent
as $\hat{h}$ never cross the threshold (dotted horizontal line) on the motile side. 
%As the zeroth order solution becomes exact in the limit
%$\hat\rho_s\rightarrow\infty$ similar profiles are found for larger
%$\hat \rho_s$. 

% The analytical solution described above is exact in the limit
% $\hat\rho_s\rightarrow\infty$.  As can be seen from Eq.~(\ref{eq:6}),
% a smaller $\hat\rho_s$ bends the cell density profile downward in
% region I. It also tends to shift the peak in cell density in region II
% to a larger $\hat z$ value due to a reduced negative curvature at the
% peak as required by Eq.~(\ref{eq:4_new}). Nevertheless, the dependence
% of the profiles on $\hat\rho_s$ close to the moving front at $\hat
% z=0$ is quite weak.

To test our analytical solution we have calculated the steady
traveling profiles by integrating Eqs.~(\ref{eq:1})-(\ref{eq:2})
numerically in the moving frame for the above boundary conditions
(thick solid lines).  As is evident from Fig.~2, the zeroth order
solution already captures the key features of the solution while the
first order solution shows quantitatively excellent agreement even at
$\hat\rho_s=4$. 

Given this good agreement, we can now use the analytical expressions to
find the stability limit of the traveling wave solution, i.e., parameter
values for which the peak height $\hat h_m$ of $\hat h_{\rm II}(\hat z)$ 
reaches the threshold value $\hat\Kh=1$. 
Let us first consider $\hat D_h=D_h/D_\rho\simeq 1$ as in the experiments.
The Green's function (\ref{G_h}) decays at a rate of order one in scaled units
when the scaled AHL diffusion length $\hat{L}_h=\sqrt{\hat D_h/\hat\beta}\sim1=\hat L_\rho$,
but much faster when $\hat{L}_h\ll 1$. In the latter case, the AHL profile follows
closely the cell density profile, reaching its peak value at the tip of the bulge.
A straightforward exercise based on Eq.~(\ref{h-linear}) of the linear case shows
$\hat h_m=\hat\rho_m=2\hat\rho_c/e$ while $\hat h_{\rm}(0)=\hat\rho_c/2=1$. Hence
$\hat h_m=4/e\simeq 1.47>\hat\Kh$. In this case the traveling wave solution is not
self-consistent. An increase of $\hat L_h$ allows immotile cells to contribute 
more to the AHL level in the motile region. Consequently $\hat h_{\rm II}(\hat z)$
flattens while $\hat\rho_c$ decreases at the same time. Eventually $\hat h_m$ drops
to a value below $\hat K_h$ to restore self-consistency of the traveling wave solution.

The actual stability limit 
can be obtained by numerically solving the
equations $\hat h_{\rm II}(\hat z_m)\equiv \hat h_m=1$ and $\partial
_{\hat z} h_{\rm II}(\hat z_m)=0$, where $\hat z_m$ is the peak position
of the AHL profile.
Using the respective analytical
profiles, we obtain the zeroth order $\hat \beta=\phi^{\rm lin}(\hat
D_h)$ (thin solid line) and first order $\hat\beta=\phi^{(1)}(\hat
D_h)$ (dashed line) phase boundaries as shown in Fig.~\ref{fig:2}.  In
the latter case, the first order AHL profile allows us to compute the
shift $\delta\hat\beta=-\varepsilon\psi(\hat D_h)\phi^{\rm lin}(\hat
D_h)$ in $\hat\beta$ that satisfies these equations to order $\varepsilon$
at a given $\hat D_h$.  
The modified boundary is then obtained from $\phi^{(1)}(\hat
D_h)=\phi^{\rm lin}(\hat D_h)\exp[-\varepsilon\psi(\hat D_h)]$.  The
function $\psi(\hat D_h)$ is given by the dotted line in the inset of
Fig. 3.  As a comparison, we have also computed the phase boundary
$\hat\beta=\phi(\hat D_h)$ where $\hat h_m=1$ using the numerically
exact traveling wave solution (thick solid line in Fig. 3).  The
agreement with the first order phase boundary $\phi^{(1)}(\hat D_h)$
is very good.

\begin{figure}[t]
  \centering
  \includegraphics[width=.43\textwidth,clip]{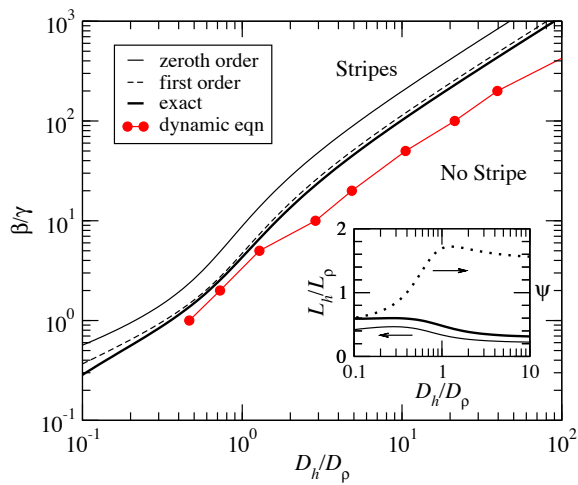}
  \caption{ Phase diagram for stripe formation. The thin solid and dashed curves are, respectively, the phase
    boundaries as calculated from our analytical solution in zeroth
    and first order in $\varepsilon=\rho_c/\rho_s$. The thick solid line
    is obtained from the numerical
    solution of Eqs.~(\ref{eq:1})-(\ref{eq:2}) in the moving frame. The red line represents
    the boundary determined from the onset of stripe patterns based on
    simulation of the full kinetic model
    (\ref{eq:1})-(\ref{eq:2}).  Inset: thin and thick solid lines give
    $L_h/L_\rho$ on the zeroth order and exact phase
    boundaries, respectively.
    The dotted line shows the function $\psi(D_h/D_\rho)$
    from the first order correction in $\varepsilon$ to the boundary
    position.
    \label{fig:2}}
\end{figure}

%The saturation effect introduced by a finite $\hat\rho_s$ on bacterial growth
%is seen to only slightly reduce the stability of the no-stripe phase.
%Qualitatively, one may argue that this is due to a somewhat broadened
%cell-density profile as seen in Fig.~\ref{fig:1} which leads to a
%larger threshold value for the AHL diffusion length $L_h$ at the
%boundary, as indicated by the thin and thick solid curves in the inset
%of Fig.~\ref{fig:2}. 

As a confirmation that our ansatz indeed captures
the dynamic instability behind the stripe formation process, we also
show in Fig.~3 (red dots) the actual onset of stripes observed from a
numerical simulation of Eqs.~(\ref{eq:1})-(\ref{eq:2}). Due to the
time it takes for transient stripes to dissipate close to the
transition with the setup of Fig. ~\ref{fig:c1}(b), the simulation tends to 
underestimate the no-stripe region. 
Thus the true phase boundary in the long-time limit is
expected to be somewhat above the red line.

This study has led to the following picture of 
the stripe formation process: the growth and lateral expansion of the
colony into an initially cell free region is described by a traveling 
wave solution. In the steadily propagating case,
the density-coupled cell motility control breaks the colony
into an immotile region behind a moving boundary and a density bulge of motile cells ahead of it.
Maximum cell density in the motile region is reached at a distance $L_\rho=\sqrt{D_\rho/\gamma}$
from the boundary.
In the experiments of Ref.~\cite{exp_paper}, the density coupling is implemented
via a small molecule AHL which provides information on cell density within a
distance $L_h=\sqrt{D_h/\beta}$. We have shown that the steadily propagating wave
is stable when $L_h$ is greater than or comparable to $L_\rho$.
In the opposite case $L_h\ll L_\rho$, instability develops as the maximum
AHL concentration in the motile region would exceed the threshold $\Kh$
for motility suppression. Instead, the colony expands with periodic nucleation of 
new immotile regions within the motile bulge ahead of the previously
formed high-density strip. Cell density behind the moving front continue to grow until
nutrient exhaustion, where the density modulation becomes frozen.
From the inset of Fig.~3 
we see that the ratio $L_h/L_\rho$ generally lies around 0.5
on the phase boundary between the two regimes.

In our system the propagating front thus drives sequential stripe
formation in an open geometry. This is very different from the
classical Swift-Hohenberg \cite{cros93} mechanism where
finite-wavelength symmetry breaking instability develops in the
bulk. The highly nonlinear and localized process in the nucleation of
new stripes also makes our mechanism different from that of pattern
formation driven by fronts propagating into a bistable system where
modulations arise during the linear instability development at the
front \cite{dee88}.  In this respect, there are some similarities
between our system and the nonperiodic Liesegang patterns since in
both cases new ``phase'' precipitates when certain critical density is
reached \cite{droz00}.  On the other hand, in the chemical systems
that exhibit Liesegang patterns, reactant density increases via
transport instead of growth.
%Furthermore, our aggregation mechanism has some
%resemblance with that of diffusion limited aggregation
%\cite{witt81}. The main difference is that in our system 
%transport not only occurs by diffusion but also by growth giving rise
%to regular patterns that form sequentially and with constant spacing.

We thank J. Tailleur and H. Levine for discussions.  The work is
supported in part by the Research Grants Council of the HKSAR under
grants HKU1/CRF/10 (JDH) and 201910 (LHT).

%\begin{appendix}

%\bibliography{prl}
% \begin{thebibliography}{99999}
% \end{thebibliography}

%\end{appendix}

%\bibitem[{\citenamefont{Witten and Sander}(1981)}]{witt81}
%\bibinfo{author}{\bibfnamefont{T.A.} \bibnamefont{Witten}} \bibnamefont{and}
%  \bibinfo{author}{\bibfnamefont{L.M.} \bibnamefont{Sander}},
%  \bibinfo{journal}{Phys. Rev. Lett.} \textbf{\bibinfo{volume}{47}},
%  \bibinfo{pages}{1400} (\bibinfo{year}{1981}).

%\begin{appendix}
% \end{appendix}

\vspace*{22cm}

%\newpage
\pagebreak[4]

\onecolumngrid
\renewcommand{\theequation}{S\arabic{equation}}
\setcounter {equation} {0}
\renewcommand{\thefigure}{S\arabic{figure}}
\setcounter {figure} {0}

\begin{center}
{\Large \bf Supporting Material: Stripe formation in bacterial systems with density-dependent
  motility}
\end{center}

As mentioned our model can be derived by coarse-graining of the
microscopic dynamics. Generally, this procedure yields a mixture of
two terms $\partial _x (\mu(h)\partial _x \rho)+\theta \partial_x
(\rho \partial _x\mu(h))$, where $\theta$ depends on the underlying
microscopic dynamics, e.g. $\theta=1$ for Ito-dynamics and
$\theta=1/2$ for Stratonovich dynamics  \cite{kampen}. With this
general coupling our model becomes
\begin{eqnarray}
\frac{\partial h}{\partial t} & = & D_h\frac{\partial^2h}{\partial
x^2}+\alpha\rho-\beta h ,\label{eq:S1}\\
\frac{\partial \rho}{\partial t} & = &  \frac{\partial^2}{\partial
x^2} \left ( \mu(h)\rho \right ) -(1-\theta)\frac{\partial}{\partial
x} \left ( \rho \frac{\partial \mu(h)}{\partial x} \right )+\gamma_0\rho \left (1-\frac{\rho}{\rho_s}\right).\label{eq:S2}
\end{eqnarray}

In the main text, we study the case $\theta=1$. In the following, we
demonstrate that for $\theta<1$ the moving front still acts as an
absorbing boundary and that our main conclusions remain
valid in this case.

{\bf Boundary conditions at the moving front $z=0$.}

The field $h(x)$ is continuous at $h=K_h$ and even
differentiable. This can be seen by integrating Eq.~(\ref{eq:S1})
from $z=0^-$ to $z=0^+$ (as in the main text $z=x-c t$) and by using $\partial h/\partial
t=-c \partial h/\partial z$.
\begin{equation}
cw=-c[h(0^+)-h(0^-)]=D_h\left. \partial h/\partial
z\right | ^{z=0^+}_{z=0^-}, \label{eq:S3}
\end{equation}
implying $\partial h(z)/\partial  z|_{0^+}=\partial h(z)/\partial z|_{0^-}$ as
$w\rightarrow 0$.

In contrast, the density $\rho(z)$ is discontinuous at $z=0$ with
\begin{eqnarray}
\rho(0^+)-\rho(0^-) & = &-\left
  . \frac{1}{c}\frac{\partial(\mu(h)\rho)}{\partial z} \right
|^{z=0^+}_{z=0^-}+\left . \frac{(1-\theta)\rho}{c} \frac{\partial
    (\mu(h))}{\partial z} \right |^{z=0^+}_{z=0^-}=-\theta\left
  . \frac{\rho}{c}\frac{\partial(\mu(h))}{\partial z} \right
|^{z=0^+}_{z=0^-} \nonumber \\ &= &\left . -\theta\frac{D_\rho\rho(0^+)}{w}\frac{\partial
    h}{\partial z}\right |_{z=0},
\label{s4}
\end{eqnarray}

where we have used that $\mu(h(z=0^+))=\mu(h(z=0^-))$. Eq.~(\ref{s4})
thus implies $\rho(0^+)=0$ as $w\rightarrow 0$. Thus, independent of
the value of $\theta$ the moving front acts as absorbing
boundary. From this analysis it is also evident that the boundary
condition $\rho(0^+)=0$ does not depend on our assumptions on the
specific functional form of the interpolating function $D_{\rho}(h)$
for $K_h-w <h<K_h$.

Finally, the slope of $\rho(0^+)$ can be determined
by integrating Eq.~(\ref{eq:S1}) from $z=0^-$ to $z=0^+$
\begin{equation}
  \label{eq:5}
\left.  \partial \rho/\partial z\right |_{z=0^+}=c \rho_c/D_{\rho},
\end{equation}
where (as in the main text) $\rho_c = \rho(0^-)$.

{\bf Results for $\theta<1$. } We have calculated the cellular density
profile and the phase diagram for different values $\theta<1$. As can
be seen from Fig.~\ref{fig:S1} the additional term in
Eq.~(\ref{eq:S2}) only leads to very small (hardly visible)
modifications of the cellular density and AHL concentration profile.
Consequently, the phase boundary is also not affected by the value of
$\theta$.

\begin{figure}[t]
  \centering
  \includegraphics[width=.93\textwidth,clip]{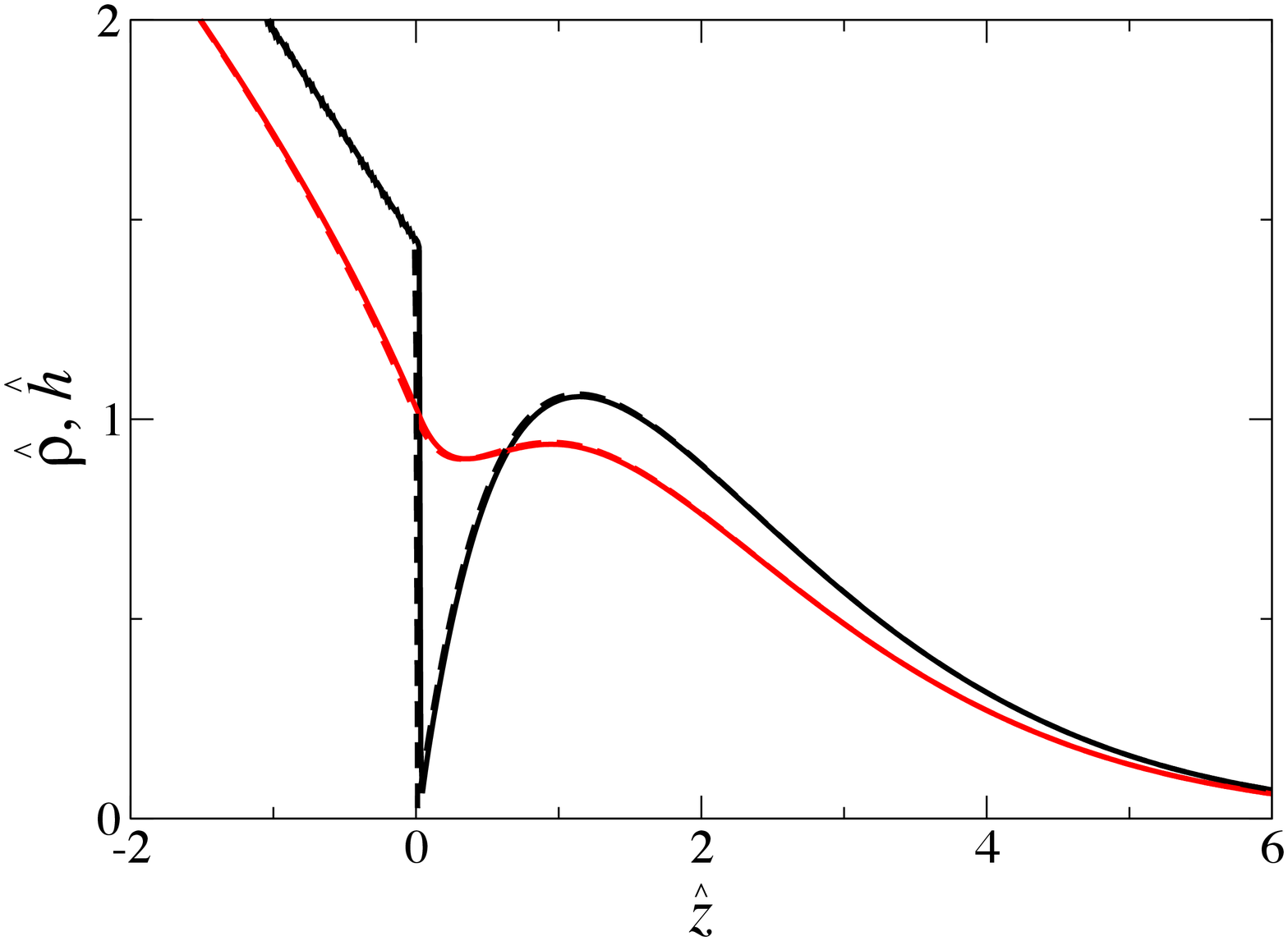}
  \caption{Profiles of scaled cellular density
    $\hat\rho(\hat z)$ (black) and AHL concentration $\hat h(\hat z)$
    (red) close to the moving front at $\hat z=0$ as calculated
    numerically from Eqs.~(\ref{eq:S1}) and (\ref{eq:S2}). The solid
    lines are for $\theta=1/2$ the dashed lines for $\theta=1$. Data are for  $\hat D_h =
    D_h/D_\rho=1.3$,
    $\hat \beta=\beta/\gamma=5$, and $\hat \rho_s=\alpha\rho_s/\beta \Kh=4$.
    \label{fig:S1}}
\end{figure}

\begin{appendix}

%\bibliography{prl}

\end{appendix}

\end{document}